\documentclass{emulateapj}

\newcommand{\br}{{\bf r}}
\newcommand{\bk}{{\bf k}}
\newcommand{\Lya}{Ly$\alpha$~}

\newcommand{\td}{{\tilde{\delta}}}

\newcommand{\etalb}{et al.}
\newcommand{\beq}{\begin{equation}}
\newcommand{\beqa}{\begin{eqnarray}}
\newcommand{\eeq}{\end{equation}}
\newcommand{\eeqa}{\end{eqnarray}}

\usepackage{apjfonts}

\begin{document}

\title{A Method for Separating the Physics from the Astrophysics of 
High-Redshift 21cm Fluctuations}

\author{Rennan Barkana\altaffilmark{1} \& Abraham
  Loeb\altaffilmark{2}}

\altaffiltext{1} {School of Physics and Astronomy, The Raymond and 
Beverly Sackler Faculty of Exact Sciences, Tel Aviv University, Tel
Aviv 69978, ISRAEL; barkana@wise.tau.ac.il}

\altaffiltext{2} {Astronomy Department, Harvard University, 60 
Garden Street, Cambridge, MA 02138; aloeb@cfa.harvard.edu}

\begin{abstract}
Fluctuations in the 21cm brightness from cosmic hydrogen at redshifts
$z\ga 6$ were sourced by the primordial density perturbations from
inflation as well as by the radiation from galaxies. We propose a
method to separate these components based on the angular dependence of
the 21cm fluctuation power spectrum. Peculiar velocities increase the
power spectrum by a factor of $\sim 2$ compared to density
fluctuations alone, and introduce an angular dependence in Fourier
space. The resulting angular structure relative to the line of sight
facilitates a simple separation of the power spectrum into several
components, permitting an unambiguous determination of the primordial
power spectrum of density fluctuations, and of the detailed properties
of all astrophysical sources of 21cm fluctuations. We also demonstrate
that there is no significant information to be gained from 21cm
measurements on large angular scales. Thus, observational analyses can
be confined to small angular areas where projection effects are
negligible, and theoretical predictions should focus on the
three-dimensional power spectrum of 21cm fluctuations.
\end{abstract}


\section{Introduction}

Following the recombination of protons and electrons less than a
million years after the big bang, the universe was filled with neutral
hydrogen (\ion{H}{1}). Hundreds of millions of years later, the first
galaxies began to reionize the cosmic gas \citep{BL01}. The spectra of
the farthest quasars indicate that reionization completed at a
redshift $z\sim 6$, a billion years after the big bang \citep{Fan,
White, Wyithe}. The hyperfine spin-flip transition of \ion{H}{1} at a
wavelength of 21 cm provides the most promising tracer of the cosmic
gas before the end of reionization. Several groups are currently
constructing low-frequency radio arrays capable of detecting the
diffuse 21cm radiation (http://space.mit.edu/eor-workshop/). Since
this radiation interacts with \ion{H}{1} through a resonant
transition, any observed wavelength selects a redshift slice of the
universe. Future measurements of the redshifted 21cm brightness as a
function of wavelength and direction should provide a
three-dimensional map of the cosmic \ion{H}{1} \citep{Hogan, Madau}.
Fluctuations in the 21cm brightness are sourced by primordial density
inhomogeneities on all scales down to the cosmological Jeans mass,
making 21cm the richest cosmological data set on the sky
\citep{Loeb04}.

The 21cm signal can be seen from epochs during which the gas was
largely neutral and deviated from thermal equilibrium with the cosmic
microwave background (CMB). The signal vanished at redshifts $z\ga
200$, when the residual fraction of free electrons after cosmological
recombination kept the gas kinetic temperature, $T_{k}$, close to the
CMB temperature, $T_\gamma$. But during $200\ga z\ga 30$ the gas
cooled adiabatically and atomic collisions kept the spin temperature
of the hyperfine level population below $T_\gamma$, so that the gas
appeared in absorption \citep{Scott,Loeb04}. As the Hubble expansion
continued to rarefy the gas, radiative coupling of $T_s$ to $T_\gamma$
began to dominate and the 21cm signal faded. When the first galaxies
formed, the UV photons they produced between the Ly$\alpha$ and Lyman
limit wavelengths propagated freely through the universe, redshifted
into the Ly$\alpha$ resonance, and coupled $T_s$ and $T_{k}$ once
again through the Wouthuysen-Field \citep{Wout,Field} effect by which
the two hyperfine states are mixed through the absorption and
re-emission of a Ly$\alpha$ photon \citep{Madau, Ciardi}. Emission of
UV photons above the Lyman limit by the same galaxies initiated the
process of reionization by creating ionized bubbles in the neutral
cosmic gas, while X-ray photons propagated farther and heated $T_{k}$
above $T_\gamma$ throughout the universe. Once $T_s$ grew larger than
$T_\gamma$, the gas appeared in 21cm emission. The ionized bubbles
imprinted a knee in the power spectrum of 21cm fluctuations
\citep{Zalda04}, which traced the \ion{H}{1} topology until the
process of reionization was completed \citep{Fur04}.

The various effects that determine the 21cm fluctuations can be
separated into two classes. On the one hand, the density power
spectrum probes basic cosmological parameters and inflationary initial
conditions, and can be calculated exactly in linear theory. On the
other hand, the radiation from galaxies, both Ly$\alpha$ radiation and
ionizing photons, involves the complex, non-linear physics of galaxy
formation and star formation. If only the sum of all fluctuations
could be measured, then it would be difficult to extract the separate
sources, and in particular, the extraction of the power spectrum would
be subject to systematic errors involving the properties of
galaxies. We show in this {\it Letter} that the unique
three-dimensional properties of 21cm measurements permit a separation
of these distinct effects. Thus, 21cm fluctuations can probe
astrophysical (radiative) sources during the epoch of first galaxies,
while at the same time separately probing physical (inflationary)
sources. In order to affect this separation most easily, it is
necessary to measure the three-dimensional power spectrum of 21cm
fluctuations, while full-sky observations are normally restricted to
angular projections. However, we show that unlike the analogous case
of CMB fluctuations, 21cm measurements on large angular scales are not
useful. Thus, observational analyses can be confined to small angular
areas where projection effects are negligible.

\section{Spin temperature history}

The spin temperature $T_s$ is defined through the ratio between the
number densities of hydrogen atoms in the excited and ground state
levels, ${n_1/ n_0}=(g_1/ g_0)\exp\left\{-{T_\star/ T_s}\right\},$
where subscripts $1$ and $0$ correspond to the excited and ground
state levels of the 21cm transition, $(g_1/g_0)=3$ is the ratio of the
spin degeneracy factors of the levels, and $T_\star=0.0682$K
corresponds to the energy difference between the levels. As long as
$T_s$ is smaller than the CMB temperature $T_{\gamma} = 2.725 (1+z)$
K, hydrogen atoms absorb the CMB, whereas if $T_s > T_{\gamma}$ they
emit excess flux. In general, the resonant 21cm interaction changes
the brightness temperature of the CMB by \citep{Scott, Madau} $T_b
=\tau \left( T_s-T_{\gamma}\right)/(1+z)$, where the optical depth at
a wavelength $\lambda=21$cm is
\beq \label{tau} 
\tau= \frac {3c\lambda^2h A_{10}n_{\rm H}} {32 \pi
k_B T_s\, (1+z)\, (dv_r/dr)} x_{\rm HI}\ ,
\eeq 
where $n_H$ is the number density of hydrogen, $A_{10}=2.85\times
10^{-15}~{\rm s^{-1}}$ is the spontaneous emission coefficient,
$x_{\rm HI}$ is the neutral hydrogen fraction, and $dv_r/dr$ is the
gradient of the radial velocity along the line of sight with $v_r$
being the physical radial velocity and $r$ the comoving distance; on
average $dv_r/dr = H(z)/ (1+z)$ where $H$ is the Hubble parameter.
The velocity gradient term arises since the 21cm scattering
cross-section has a fixed thermal width which translates through the
redshift factor $(1+v_r/c)$ to a fixed interval in velocity
\citep{Sobolev}.

For the concordance set of cosmological parameters \citep{WMAP}, the
mean brightness temperature on the sky at redshift $z$ is $T_b = 28\,
{\rm mK}\,
[({1+z})/{10}]^{1/2} \left[({T_s - T_{\gamma}})/{T_s}\right]
\bar{x}_{\rm HI}$, where $\bar{x}_{\rm HI}$ is the mean neutral
fraction of hydrogen.  The spin temperature itself is coupled to $T_k$
through the spin-flip transition, which can be excited by collisions
or by the absorption of \Lya photons.  As a result, the combination
that appears in $T_b$ becomes \citep{Field} $(T_s - T_{\gamma})/T_s =
[x_{\rm tot}/(1+ x_{\rm tot})] \left(1 - T_{\gamma}/T_k \right)$,
where $x_{\rm tot} = x_{\alpha} + x_c$ is the sum of the radiative and
collisional threshold parameters. These parameters are $x_{\alpha} =
{4 P_{\alpha} T_\star}/{27 A_{10} T_{\gamma}}$ and $x_c = {4
\kappa_{1-0}(T_k)\, n_H T_\star}/{3 A_{10} T_{\gamma}}$, where
$P_{\alpha}$ is the \Lya scattering rate which is proportional to the
\Lya intensity, and $\kappa_{1-0}$ is tabulated as a function of $T_k$
\citep{AD, Zyg}. The coupling of the spin temperature
to the gas temperature becomes substantial when $x_{\rm tot} \ga 1$.

\section{Brightness temperature fluctuations}

Although the mean 21cm emission or absorption is difficult to measure
due to bright foregrounds, the unique character of the fluctuations in
$T_b$ allows for a much easier extraction of the signal
\citep{Shaver, Zalda04, Miguel1, Miguel2, Santos}. In general, 
the fluctuations in $T_b$ can be sourced by fluctuations in gas
density, temperature, neutral fraction, radial velocity gradient, and
\Lya flux (through $x_{\alpha}$). Adopting the notation $\delta_A$ for
the fractional fluctuation in quantity $A$ (with a lone $\delta$
denoting density perturbations), we find
\beqa \label{dsignal} \delta_{T_b} & = & \left( 1 +
\frac{x_c} {\tilde{x}_{\rm tot}} \right) \delta + \frac{x_{\alpha}}
{\tilde{x}_{\rm tot}} \delta_{x_{\alpha}} + \delta_{x_{\rm HI}} -
\delta_{d_rv_r} \nonumber \\ & & + (\gamma_a - 1) \left[ \frac{T_{\gamma}}
{T_k - T_{\gamma}} + \frac{x_c} {\tilde{x}_{ \rm tot}}\, \frac{d
\log(\kappa_{1-0})} {d \log(T_k)} \right]\, \delta \ , \eeqa where the 
adiabatic index is $\gamma_a = 1 + (\delta_{T_k} / \delta)$, and we
define $\tilde{x}_{\rm tot} \equiv (1 + x_{\rm tot}) x_{\rm
tot}$. Taking the Fourier transform, we obtain the power spectrum of
each quantity; e.g., the total power spectrum $P_{T_b}$ is defined by
\beq \label{pTb} \langle \td_{T_b} (\bk_1) \td_{T_b} (\bk_2) \rangle = 
(2\pi)^3 \delta^D(\bk_1+\bk_2) P_{T_b}(\bk_1)\ , \eeq where $\td_{T_b}
(\bk)$ is the Fourier transform of $\delta_{T_b}$, $\bk$ is the comoving
wavevector, and $\langle \cdots \rangle$ denotes an ensemble average.

\section{The separation of powers}

The fluctuation $\delta_{T_b}$ consists of a number of isotropic
sources of fluctuations plus the peculiar velocity term
$-\delta_{d_rv_r}$. Its Fourier transform is simply proportional to
that of the density field \citep{kaiser, Indian}, \beq \label{vgrad}
\td_{d_rv_r} = -\mu^2 \td , \eeq where $\mu = \cos\theta_k$ in terms 
of the angle $\theta_k$ of $\bk$ with respect to the line of sight. The
$\mu^2$ dependence in this equation results from taking the radial
component ($\propto \mu$) of the peculiar velocity, and then the
radial component ($\propto \mu$) of its gradient. Intuitively, a
high-density region possesses a velocity infall towards the density
peak, implying that a photon must travel further from the peak in
order to reach a fixed relative redshift, compared with the case of
pure Hubble expansion. Thus the optical depth is always increased by
this effect in regions with $\delta > 0$. This phenomenon is most
properly termed {\it velocity compression}.

We therefore write the fluctuation in Fourier space as 
\beq \label{Tbk} \td_{T_b} (\bk) = \mu^2 \td(\bk) + \beta \td(\bk) + 
\td_{\rm rad}(\bk)\ , \eeq
where we have defined a coefficient $\beta$ by collecting all terms
$\propto \delta$ in equation~(\ref{dsignal}), and have also combined
the terms that depend on the radiation fields of \Lya photons and
ionizing photons, respectively. We assume that these radiation fields
produce isotropic power spectra, since the physical processes that
determine them have no preferred direction in space. The total power
spectrum is
\beqa \label{powTb}
P_{T_b}(\bk) & = & \mu^4 P_{\delta}(k) + 2 \mu^2 [\beta P_{\delta}(k) +
P_{\delta-{\rm rad}}(k)]+ \nonumber \\ & & [\beta^2 P_{\delta}(k) +
P_{\rm rad}(k) + 2 \beta P_{\delta-{\rm rad}}(k)]\ , 
\eeqa 
where we have defined the power spectrum $P_{\delta-{\rm
rad}}$ as the Fourier transform of the cross-correlation 
function,
\beq \label{xi} 
\xi_{\delta-{\rm rad}} (r) =
\langle \delta(\br_1)\, \delta_{\rm rad} (\br_1 + \br) \rangle\ .  
\eeq 

We note that a similar anisotropy in the power spectrum has been
previously derived in a different context, that of measuring density
fluctuations from galaxy redshift surveys. In that case, the use of
redshifts to estimate distances changes the apparent density of
galaxies along the line of sight
\citep{kaiser,lilje,hamilton,fisher}. However, in the case of galaxies
there is no analog to the method that we demonstrate below for
separating in 21cm fluctuations the effect of initial conditions from
that of later astrophysical processes.

We may now calculate the angular power spectrum of the brightness
temperature on the sky at a given redshift -- corresponding to a
comoving distance $r_0$ along the line of sight. The brightness
fluctuations can be expanded in spherical harmonics with expansion
coefficients $a_{lm}(\nu)$, where the angular power spectrum
\beqa \label{cldef} C_{l}(r_0) & = & \langle |a_{lm}(\nu)|^2 \rangle 
=4 \pi \int \frac {d^3k} {(2\pi)^3} \biggl[ G_l^2(k r_0) P_{\delta}(k)
+ \nonumber \\ & & 2 P_{\delta-{\rm rad}}(k) G_l(k r_0) j_l(k r_0) +
P_{\rm rad}(k) j_l^2(k r_0) \biggl] \ , \eeqa with $G_l(x)
\equiv J_l(x) + (\beta - 1) j_l(x)$ and $J_l(x)$ being a linear 
combination of spherical Bessel functions \citep{Indian}.

The velocity gradient term has been previously neglected in 21cm
calculations, except for its effect on the sky-averaged power and on
radio visibilities \citep{tozzi, Indian, Indian2}. In the simple case
where $\beta = 1$ and only the density and velocity terms contribute,
the velocity term increases the total power by a factor of $\langle
(1+\mu^2)^2 \rangle = 1.87$ in the spherical average. However, instead
of averaging the signal, we can use the angular structure of the power
spectrum to greatly increase the discriminatory power of 21cm
observations. We may break up each spherical shell in $\bk$ space into
rings of constant $\mu$ and construct the observed
$P_{T_b}(k,\mu)$. Considering equation~(\ref{powTb}) as a polynomial
in $\mu$, i.e., $\mu^4 P_{\mu^4} + \mu^2 P_{\mu^2} + P_{\mu^0}$, we
see that the power at just three values of $\mu$ is required in order
to separate out the coefficients of 1, $\mu^2$, and $\mu^4$ for each
$k$.

If the velocity compression were not present, then only the
$\mu$-independent term (times $T_b^2$) would be observed, and its
separation into the five components ($T_b$, $\beta$, and three power
spectra) would be difficult and subject to degeneracies. Once the power has
been separated into three parts, however, the $\mu^4$ coefficient can be
used to measure the density power spectrum directly, with no interference
from any other source of fluctuations. Since the overall amplitude of the
power spectrum, and its scaling with redshift, are well determined from the
combination of the CMB temperature fluctuations and galaxy surveys, the
amplitude of $P_{\mu^4}$ directly determines the mean brightness
temperature $T_b$ on the sky, which measures a combination of $T_s$ and
$\bar{x}_{\rm HI}$ at the observed redshift. Once $P_{\delta}(k)$ has been
determined, the coefficients of the $\mu^2$ term and the $\mu$-independent
term must be used to determine the remaining unknowns, $\beta$,
$P_{\delta-{\rm rad}}(k)$, and $P_{\rm rad}(k)$. Since the coefficient
$\beta$ is independent of $k$, determining it and thus breaking the last
remaining degeneracy requires only a weak additional assumption on the
behavior of the power spectra, such as their asymptotic behavior at large
or small scales. If the measurements cover $N_k$ values of wavenumber $k$,
then one wishes to determine $2 N_k + 1$ quantities based on $2 N_k$
measurements, which should not cause significant degeneracies when $N_k \gg
1$. Even without knowing $\beta$, one can probe whether some sources of
$P_{\rm rad}(k)$ are uncorrelated with $\delta$; the quantity $P_{\rm
un-\delta}(k) \equiv P_{\mu^0}- P_{\mu^2}^2/(4 P_{\mu^4})$ equals $P_{\rm
rad} - P_{\delta-{\rm rad}}^2 / P_{\delta}$, which receives no contribution
from any source that is a linear functional of the density distribution.

\section{Specific epochs}

At $z \sim 35$, galaxy formation is in its infancy, while collisions
are effective due to the high gas density \citep{Loeb04}. Thus, $x_{\rm
tot} = x_c \gg x_{\alpha}$, $x_{\rm HI} = 1$, and the gas cools
adiabatically (with $\gamma_a = 5/3$). In addition to measuring the
density power spectrum\citep{Loeb04}, one can also trace the redshift
evolution of $n_{\rm HI}$, $T_{\gamma}$, and $T_k$, and verify that
collisions are causing the fluctuations.

At $z\la 35$, collisions become ineffective but the first stars
produce a cosmic background of \Lya photons that couples $T_s$ to
$T_k$. During the period of initial \Lya coupling, fluctuations in the
\Lya flux translate into fluctuations in the 21cm brightness
\citep{BL04}. This signal can be observed from $z \sim 25$ until the
\Lya coupling is completed (i.e., $x_{\rm tot} \gg 1$) at $z \sim
15$. At a given redshift, each atom sees
\Lya photons that were originally emitted at earlier times at rest-frame 
wavelengths between \Lya and the Lyman limit. Distant sources are time
retarded, and since there are fewer galaxies in the distant, earlier
universe, each atom sees sources only out to an apparent source
horizon of $\sim 100$ comoving Mpc at $z \sim 20$. A significant
portion of the flux comes from nearby sources, because of the $1/r^2$
decline of flux with distance, and since higher Lyman series photons,
which are degraded to \Lya photons through scattering, can only be
seen from a small redshift interval that corresponds to the wavelength
interval between two consecutive atomic levels.

\begin{figure}
\plotone{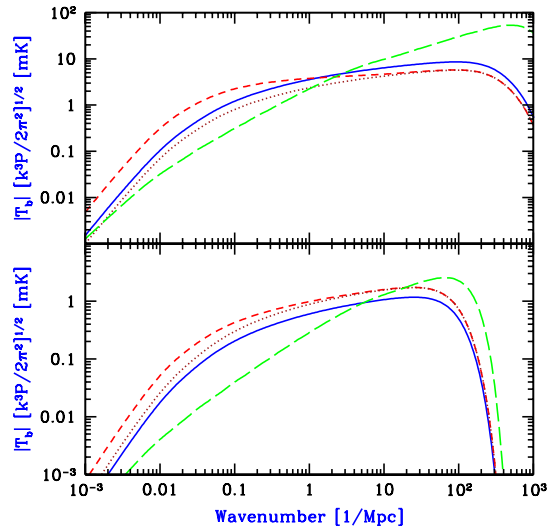}
\caption{Observable power spectra during the period of initial \Lya 
coupling. {\it Upper panel:} Assumes adiabatic cooling. {\it Lower
panel:} Assumes pre-heating to 500 K by X-ray sources. We show
$P_{\mu^4}=P_{\delta}$ (solid curves), $P_{\mu^2}$ (short-dashed
curves), and $P_{\rm un-\delta}$ (long-dashed curves). We also show
for comparison $2 \beta P_{\delta}$ (dotted curves).}
\label{Pkfig} 
\end{figure}

There are two separate sources of fluctuations in the \Lya flux
\citep{BL04}. The first is density inhomogeneities.  Since gravitational
instability proceeds faster in overdense regions, the biased
distribution of rare galactic halos fluctuates much more than the
global dark matter density.  When the number of sources seen by each
atom is relatively small, Poisson fluctuations provide a second source
of fluctuations. Unlike typical Poisson noise, these fluctuations are
correlated between gas elements at different places, since two nearby
elements see many of the same sources. Assuming a scale-invariant
spectrum of primordial density fluctuations, and that $x_{\alpha}=1$
is produced at $z=20$ by galaxies in dark matter halos where the gas
cools efficiently via atomic cooling, we show in Figure 1 the
predicted observable power spectra. The figure shows that $\beta$ can
be measured from the ratio $P_{\mu^2} / P_{\mu^4}$ at $k \ga 1$
Mpc$^{-1}$, allowing the density-induced fluctuations in flux to be
extracted from $P_{\mu^2}$, while only the Poisson fluctuations
contribute to $P_{\rm un-\delta}$. Each of these components probes the
number density of galaxies through its magnitude, and the distribution
of source distances through its shape. Measurements at $k \ga 100$
Mpc$^{-1}$ can independently probe $T_k$ because of the smoothing
effects of the gas pressure and the thermal width of the 21cm line.

After \Lya coupling and X-ray heating are both completed, reionization
continues. Since $\beta = 1$ and $T_k \gg T_{\gamma}$, the
normalization of $P_{\mu^4}$ directly measures the mean neutral
hydrogen fraction, and one can separately probe the density
fluctuations, the neutral hydrogen fluctuations, and their
cross-correlation.

\section{Fluctuations on large angular scales}

Full-sky observations must normally be analyzed with an angular and
radial transform, rather than a Fourier transform which is simpler and
yields more directly the underlying 3D power spectrum. In an angular
transform on the sky, an angle of $\theta$ radians translates to a
spherical multipole $l \sim 3.5/ \theta$. For measurements on a screen
at a comoving distance $r_0$, a multipole $l$ normally measures 3D
power on a scale of $k^{-1} \sim \theta r_0 \sim 35/l$ Gpc for $l\gg
1$, since $r_0 \sim 10$ Gpc at $z \ga 10$. This estimate fails at $l
\la 100$, however, when we consider the sources of 21cm
fluctuations. The angular projection implied in $C_l$ involves a
weighted average [equation~(\ref{cldef})] which favors large scales
when $l$ is small, but density fluctuations possess little large-scale
power, and the $C_l$ are dominated by power around the peak of $k
P_\delta(k)$, at a few tens of comoving Mpc.

\begin{figure}
\plotone{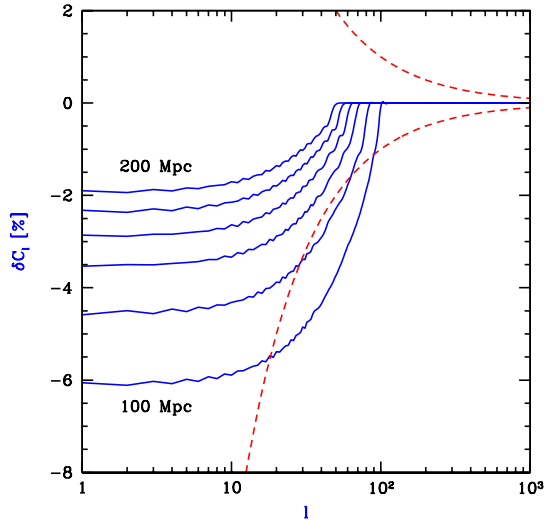}
\caption{Effect of large-scale power on the angular 
power spectrum of 21cm anisotropies on the sky. This example shows the
power from density fluctuations and velocity compression, assuming a
hot IGM at $z=12$ with $T_s=T_k \gg T_{\gamma}$. We show the $\%$
change in $C_l$ if we were to cut off the power spectrum above $1/k$
of 200, 180, 160, 140, 120, and 100 Mpc (top to bottom). Also shown
for comparison is the cosmic variance for averaging in bands of
$\Delta l \sim l$ (dashed lines).}
\label{clfig} 
\end{figure}

Figure 2 shows that for density and velocity fluctuations, even the
$l=1$ multipole is affected by power at $k^{-1} > 200$ Mpc only at the
$2\%$ level. Due to the small number of large angular modes available
on the sky, the expectation value of $C_l$ cannot be measured
precisely at small $l$. Figure 2 shows that this precludes new
information from being obtained on scales $k^{-1} \ga 130$ Mpc using
angular structure at any given redshift. An optimal observing strategy
would be to divide the sky into separate angular pixels of size
$\theta \la 2^{\circ}$, in which the curvature changes distances by
only $\theta^2/24 \sim 7 \times 10^{-5}$.  This would allow a direct
3D Fourier transform, which is also the most natural way of analyzing
radio observations \citep{Miguel1, Miguel2}.

\section{Summary}

We have argued that observers should try to measure the anisotropic
form of the power spectrum of 21cm fluctuations. Since measurements on
large angular scales are not worthwhile, the three-dimensional power
spectrum may indeed be measured by dividing the sky into small angular
areas. The anisotropy itself, which is due to the effect of peculiar
velocities, facilitates a simple separation of the power spectrum into
three separate components. One component allows an unambiguous
determination of the primordial power spectrum of gas density
fluctuations, which if measured will probe basic cosmological
parameters as well as inflation. The other two components probe the
properties of astrophysical (radiative) sources of 21cm fluctuations.

\acknowledgments

This work was supported in part by NSF grants AST-0204514, AST-0071019
and NASA grant NAG 5-13292 . R.B. is grateful for the kind hospitality
of the Harvard-Smithsonian CfA, and acknowledges the support of an
Alon Fellowship at Tel Aviv University and of Israel Science
Foundation grant 28/02/01.

\end{document}